\documentclass[a4paper,12pt]{article}

\usepackage{amsmath,amssymb,array,amsfonts}
\usepackage[dvipdfmx]{graphicx}
\usepackage{comment,xcolor}
\usepackage{cite}
\usepackage{nicefrac}
\usepackage[hang, small, bf]{caption}

\allowdisplaybreaks

\def\e {{\rm e}}	
\def\s {{\rm s}}
\def\H {{\rm H}}
\def\SM{{\rm SM}}
\def\Tr{{\rm Tr}}

\def\dis{\displaystyle}

\newcommand{\bs}{\boldsymbol}

\def\dy{\int_{-\pi R}^{\pi R}\!\!\!\!\!\!\!\!{\rm d}y\,}%

\makeatletter
    
    \@addtoreset{equation}{section}
\makeatother

\usepackage{hyperref}
\hypersetup{
			colorlinks= true,
			linkcolor= purple,%
			citecolor = cyan
			}%

\begin{document}
\setlength{\baselineskip}{18pt}
\begin{titlepage}

\begin{flushright}
KOBE-TH-12-01
\end{flushright}
\vspace{1.0cm}
\begin{center}
{\Large\bf CP Violation due to Flavor Mixing\\[5pt] in Gauge-Higgs Unification} 
\end{center}
\vspace{25mm}

\begin{center}
{\large
Yuki Adachi, 
%
Nobuaki Kurahashi$^*$, 
%
Nobuhito Maru$^{**}$
%
and Kazuya Tanabe$^*$
%
}
\end{center}
\vspace{1cm}
\centerline{{\it
Department of Sciences, Matsue College of Technology,
Matsue 690-8518, Japan.}}

\centerline{{\it
$^*$Department of Physics, Kobe University,
Kobe 657-8501, Japan.}}

\centerline{{\it
$^{**}$Department of Physics, and Research and Education Center for Natural Sciences,}}
\centerline{{\it
Keio University,
Yokohama 223-8521, Japan.
}}
%
%
\vspace{2cm}
\centerline{\large\bf Abstract}
\vspace{0.5cm}

We discuss CP violation in the five dimensional $SU(3) \otimes SU(3)_{\rm color}$ gauge-Higgs unification scenario 
in which the fifth dimension is compactified on an orbifold $S^1\!/Z_2$.
It is shown that CP-violating phase appears 
even in the two generation case 
in contrast with the fact that at least three generations are required to break CP symmetry in the Standard Model.
As our prediction, 
we obtain the  phenomenological constraint on the compactification scale 
by comparing the typical CP-violating observables, namely  $\varepsilon_K$ parameter and the mass difference of two neutral $K$ mesons $\Delta m_K$
with experimental data. 

\end{titlepage}




\newpage
\section{Introduction}

Gauge-Higgs unification (GHU) \cite{GH} opens an new avenue to go beyond the standard model (SM) 
since the hierarchy problem is solved without supersymmetry by identifying Higgs scalar field with an extra spatial component of gauge fields. 
In other words, the quantum correction to the Higgs boson mass becomes finite due to the higher dimensional gauge symmetry 
though the theory is non-renormalizable.
The finiteness of the Higgs mass has been verified in various models \cite{ABQ, MY}.%
\footnote{For the case of gravity-gauge-Higgs unification, see \cite{HLM}.}
It is interesting to find other finite observables similar to the Higgs mass in this scenario. 
In this regard, several studies were done concerning about the $S$ and $T$ parameters, 
gluon-fusion interaction and fermion anomalous magnetic moment and electric dipole moment \cite{LM, LHC, ALM1}.

On the other hand, 
the flavor mixing and CP violation in this scenario is non-trivial issue 
since the Yukawa and the gauge interaction are unified into higher dimensional gauge interaction. 
In this context, the Yukawa coupling becomes \lq\lq real" and weak gauge eigenstate is essentially equal to  mass eigenstates,
it seems to be difficult to incorporate CP violation and flavor mixing.
We have already discussed the flavor mixing in the GHU \cite{2010AKLM, 2011AKLM, 2011AKMT} 
and pointed out that 
the flavor mixing is realized by an interplay between the non-degenerate fermion bulk mass term and non-trivial brane-localized mass term.
As a result, the flavor changing neutral current (FCNC)
via non-zero Kaluza-Klein (KK) gauge boson exchange occurs at tree level even in QCD sector.

Focussing on the CP violation, 
several types of CP violation in the GHU have been already investigated \cite{ALM3, LMN}.
In the paper \cite{ALM3},
the Higgs boson  behaves as a CP odd scalar and CP symmetry breaks down spontaneously with electroweak symmetry breaking.
In the paper \cite{LMN},
a complex structure can be embedded into the compactified space and
the CP violation is obtained by incompatibility with the orbifolding parity. 
Both of the above CP violation are specific to higher dimensional gauge theory,
however, the Kobayashi-Maskawa (KM) type CP violation is not studied precisely.

In this paper we address the issue of CP violation related to flavor mixing in the GHU scenario.
It is highly non-trivial problem to explain the CP violation
in addition to the variety of fermion masses and flavor mixings in this scenario,
since Yukawa coupling is the real and universal gauge couplings.
To see how CP symmetry breaks in such models,
we need to mention the mechanism to realize flavor mixing.
In the $SU(3)\otimes SU(3)_{\rm color}$ model,
$n$ set of $\bf 3$ and $\bf{\bar 6}$ representation of $SU(3)$ should be introduced to reproduce $n$ generations of quark.%
\footnote{%
More precisely, $\bf{\overline{15}}$ representation should be introduced to embedding top quark, similar argument can be expanded.
}
Since each representation has two doublets $Q_3$ and $Q_6$, namely two massless quark doublets appear per generation, 
we identify the SM quark doublet $Q_{\rm SM}$ as follows.
\begin{align}
	\left[\begin{array}{cc}
	 U_1 & U_3\\[3pt]
	 U_2 & U_4
	\end{array}\right]\!\!
	\left[\begin{array}{c}
	 Q_{\H L}(x)\\[3pt]
	 Q_{\SM L}(x)
	\end{array}\right]
  = \left[\begin{array}{c}
	 Q_{3L}(x)\\[3pt]
	 Q_{6L}(x)
	\end{array}\right]
\end{align}
where $U_3$, $U_4$ are $n\times n$ matrices 
which indicate that the SM quark doublets are given by what kind of mixture of $Q_{3L}(x)$ and $Q_{6L}(x)$
and compose of a $2n\times 2n$ unitary matrix together with $U_1$ and $U_2$.
The eigenstate $Q_\H$ becomes massive and decouples from the low energy processes by the brane-localized mass term,
while $Q_\SM$ remains massless at this stage
and is identified with the SM quark doublet.
$U_3$ and $U_4$ should satisfy the following unitarity condition:
\begin{align}
	\label{unitary_cond} 
	U_3^\dag U_3+U_4^\dag U_4={\bf 1}_{n\times n}.  
\end{align} 

Now we discuss the counting of physical CP-violating phases.
Note that $U_3$ and $U_4$ generically have complex components,
they potentially violate CP symmetry.
These $n\times n$ complex matrices $U_3$ and $U_4$ are known to be written
in a product of an unitary matrix and a diagonal matrix:
\begin{align}
\begin{cases} 
	~U_3
  = 
	P_3
	\,\mathcal U\,
	P_3'\,
	{\rm diag}\Big(c_{a_1}, c_{a_2}, \cdots, c_{a_n}\Big) 
	\\[8pt]
	~U_4
  = 
  	P_4\,
	\mathcal V\,
	P_4'\,
  {\rm diag}\Big(s_{a_1}, s_{a_2},  \cdots, s_{a_n}\Big)
\end{cases}
\end{align}
where $s_{a_i} \equiv \sin a_i$, $c_{a_i} \equiv \cos a_i$.
In the above expression,
we use the fact that the arbitrary unitary matrices $V$ can be always decomposed into 
some phase matrices $P$ and $P'$ with $n$ and $n-1$ phases respectively : $V = PV'P'$.
Then the $\mathcal U$ and $\mathcal V$ have $\frac{(n-1)(n-2)}2$ phases, respectively.

Next, we focus on the counting of CP-violating phases in this model.
Since the phase matrices $P_3$ and $P_4$ in the left-hand side of $\mathcal U$ and $\mathcal V$ act on the right-handed up- and down-type quark,
they can be eliminated by the re-definition of singlet quarks.
Since a similar discussion on phase matrices in the right-hand side of $\mathcal U$ and $\mathcal V$ can be applied,
they can be eliminated by the re-definition of the $Q_{\rm SM}$.
However, both phases matrix denoted by $P_3'$ and $P_4'$ commonly act on the $Q_{{\rm SM}L}$
and thus only $n-1$ phases of $P_3'$ and $P_4'$ can be absorbed by the $Q_{{\rm SM}L}$.
Then the remaining physical phases are given by the sum of the phases
in the $\mathcal U$, $\mathcal V$ and $P_3'$ or $P_4'$:
\begin{align}
\frac{(n-1)(n-2)}{2}+\frac{(n-1)(n-2)}{2}+(n-1)
  = (n-1)^2 .
\end{align}
Since there are FCNC vertices in the strong interaction
as we have discussed before \cite{2010AKLM},
the CP-violating phase appears in the interactions
between the zero-mode fermion and non-zero KK gluons even in the two generation scheme in our model.
Note that the above argument is consistent with the KM theory,
the phases in the Yukawa interactions are completely removed
by the ordinary field re-definition.

This paper is organized as follows.
The next section after this introduction,
we construct the model in the two generation scheme as a simplest example of CP-violating model.
As an application of the CP violation due to the flavor mixing discussed in section 3,
we calculate the Wilson coefficient caused by the $\Delta S = 2$ process in section 4,
$K^0-\bar K^0$ mixing via non-zero KK gluon exchange at the tree level
in order to compare $\varepsilon_K$ parameter which is known as a typical CP-violating observables in our model with the experimental result.
We also obtain the lower bound for the compactification scale. 
Section 5 is devoted to our conclusions.

\section{The Model}
Although the model we consider in this paper is the same as the one taken
in \cite{2010AKLM, 2011AKLM}, we briefly describe the model for completeness.
The model taken in this paper is a five dimensional (5D)
$SU(3) \otimes SU(3)_{\rm color}$ GHU model
compactified on an orbifold $S^1\!/Z_2$ with a radius $R$ of $S^1$.
As matter fields, we introduce two generations of bulk fermion
in the ${\bf 3}$ and the $\bar{\bf6}$ dimensional representations of $SU(3)$ gauge group
denoted by a column vector and a $3\times3$ matrix,
$\psi^i({\bf3})$ and $\psi^i(\bar{\bf6})$ ($i = 1, 2$) \cite{BN}.

The bulk lagrangian is given by
\begin{align}
	\mathcal L
 =&	-\!\frac12\Tr\big(F_{MN}F^{MN}\big)-\frac12\Tr\big(G_{MN}G^{MN}\big)\notag\\*
  &	+\bar\psi^i({\bf 3})\big\{i\!\not\!\!D_3 -M_i\epsilon(y)\big\}\psi^i({\bf 3})
	+\frac12
	 \Tr\Big[
	 \bar\psi^i({\bar{\bf 6}})\big\{i\!\not \!\!D_6 -M_i\epsilon(y)\big\}\psi^i({\bar{\bf 6}})
	 \Big]
\end{align}
where
\begin{subequations}
\begin{align}
	F_{MN}
 &=	\partial_MA_N-\partial_NA_M -ig\big[A_M,A_N\big],\\
	G_{MN}
 &=	\partial_MG_N-\partial_NG_M -ig_\s\big[G_M,G_N\big],\\
	\not\!\!D_3 \psi^i({\bf 3}) 
 &= \varGamma^M(\partial_M-igA_M-ig_\s G_M)\psi^i({\bf 3}),\\
 	\not\!\!D_6 \psi^i({\bar{\bf 6}}) 
 &= \varGamma^M
	\Big[
	\partial_M\psi^i({\bar{\bf 6}})
	+ig\big\{A_M^{\ast}\psi^i({\bar{\bf 6}})+\psi^i({\bar{\bf 6}})(A_{M})^\dagger\big\}
	-ig_\s G_M\psi^i({\bar{\bf 6}})
	\Big],
\end{align}
\end{subequations}
with $G_M$ being understood to act on the color index, not explicitly written here.
The gauge fields $A_M$ and $G_M$ are written in a matrix form,
e.g. $A_M = A_M^a\frac{\lambda^a}2$ in terms of Gell-Mann matrices $\lambda^{a}$.
$M, N = 0,1,2,3,5$ and the 5D gamma matrices
are given by $\varGamma^M = \big(\gamma^\mu\,,\,i\gamma^5\big)$ ($\mu = 0,1,2,3$).
$g$ and $g_\s$ are 5D gauge coupling constants
of $SU(3)$ and $SU(3)_{\rm color}$, respectively.
$M_i$ are generation dependent bulk mass parameters of the fermions
accompanied by the sign function $\epsilon (y)$.
As was discussed in the introduction, here we take the base
where the bulk mass term is flavor-diagonal.

The periodic boundary condition is imposed along $S^1$ and $Z_2$ parity
assignments are taken for gauge fields as
\begin{subequations}
\begin{alignat}{3}
	A_\mu(-y) &= P A_\mu(y) P^{-1} &\quad, \qquad A_y(-y) &= -P A_y(y) P^{-1}\ ,\\
	G_\mu(-y) &= G_\mu(y) &\quad, \qquad G_y(-y) &= -G_y(y)
\end{alignat}
\end{subequations}
where the orbifolding matrix is defined as $P={\rm diag}(-,-,+)$
and operated in the same way at the fixed points $y = 0, \pi R$.
We can see that the gauge symmetry $SU(3)$ is explicitly broken
to $SU(2) \times U(1)$ by the boundary conditions.
The gauge fields with $Z_2$ odd parity and even parity are expanded
by use of mode functions,
\begin{align}
	S_n(y)
 &= \frac1{\sqrt{\pi R}}\sin\frac nRy
	\quad , \qquad
	C_n(y)
  = \frac1{\sqrt{2^{\delta_{n,0}} \pi R}}\cos\frac nRy\ ,
\end{align}
respectively.

A chiral theory is realized in the zero-mode sector by $Z_2$ orbifolding.
The fermions are also expanded by an ortho-normal set of mode functions.
Here we will focus on the zero-mode sector
which are necessary for the argument of flavor mixing:
\begin{subequations}
\begin{align}
	\Psi^i({\bs3})
 &\supset
	Q_{3L}^if_L^i(y) \oplus d_R^if_R^i(y)\ ,\\
	\Psi^i(\bar{\bs6})
 &\supset
	\Sigma_R^if_R^i(y) \oplus Q_{6L}^if_L^i(y) \oplus u_R^if_R^i(y)\ 
\end{align}
\end{subequations}
where the mode function for the zero mode of each chirality is given in \cite{ALM3}:
\begin{align}
	f^{i}_L(y)
 &= \sqrt{\frac{M_i}{1-\e ^{-2\pi RM_i}}}\e^{-M_i|y|},\quad
	f^{i}_R(y)
  = \sqrt{\frac{M_i}{\e^{2\pi RM_i}-1}}\e^{M_i|y|}. 
\end{align}

We notice that there are two left-handed quark doublets $Q_{3L}$ and $Q_{6L}$
per generation in the zero-mode sector,
which are massless before electroweak symmetry breaking.
In the one generation case, for instance,
one of two independent linear combinations of these doublets should correspond
to the quark doublet in the SM,
but the other one should be regarded as an exotic state.
Moreover, we have an exotic fermion $\Sigma_R$.
We therefore introduce brane localized four dimensional (4D) Weyl spinors
to form $SU(2) \times U(1)$ invariant brane localized Dirac mass terms
in order to remove these exotic massless fermions from the low-energy effective theory
\cite{BN, ACP}.

Some comments on this model are in order. 
The predicted Weinberg angle of this model is not realistic, $\sin^2 \theta_W = \frac34$. 
Possible modification is to introduce an extra $U(1)$ \cite{2011AKMT}
or the brane localized gauge kinetic term \cite{SSS}.
However, the wrong Weinberg angle does not affect our argument,
since our interest is $K^0$\,--\,$\bar K^0$ mixing via KK gluon exchange in the QCD sector,
whose amplitude is independent of the Weinberg angle.

\section{CP violation due to flavor mixing}

As was mentioned in introduction,
the CP phase remains even in the two generation scheme in our model.
For an illustrative purpose
to confirm the mechanism of CP violation due to flavor mixing,
we will see how the realistic quark masses and mixing are reproduced.
Then, $2\times2$ matrices $U_3$ and $U_4$ can be written without loss of generality as,
\begin{align} \label{parametrization2dCP}
	U_3
  = \!\left[
	\begin{array}{cc}
	 \cos\theta & -\!\sin\theta \\[3pt]
	 \sin\theta & \cos\theta
	\end{array}
	\right]\!\!
	\left[
	\begin{array}{cc}
	 c_{a_1} & 0 \\[3pt]
	 0 & c_{a_2}
	\end{array}
	\right]
	\quad , \qquad
	U_4
  = \!\left[
	\begin{array}{cc}
	 \cos\theta' & -\!\sin \theta' \\[3pt]
	 \sin\theta' & \cos\theta'
	\end{array}
	\right]\!\!
	\left[
	\begin{array}{cc}
	 s_{a_1} & 0 \\[3pt]
	 0 & s_{a_2}\e^{i\gamma}
	\end{array}
	\right]
\end{align}
where the CP-violating phases $\gamma$ do not need to appear in $U_4$
but may appear in $U_3$ if we wish.
It is a matter of convention.

In this case,
Yukawa couplings are read off from the higher dimensional gauge interaction of $A_y$,
whose zero mode is the Higgs field $H(x)$:
\begin{align}
	-\frac{g_4}2\!
	\left\{
	\left\langle H^{\dagger}\right\rangle\bar d_R^{i}(x)
	I_{RL}^{i(00)}U_3^{ij}Q_{\SM L}^{j}(x)
	+\sqrt2\left\langle H^t \right\rangle i\sigma^2
	 \bar u_R^{i}(x)I_{RL}^{i(00)}U_4^{ij}Q_{\SM L}^{j}(x)
	\right\}
	+{\rm h.c.}
\end{align}
where $g_4 \equiv \frac g{\sqrt{2\pi R}}$
and $I_{RL}^{(00)}$ is an overlap integral of mode functions of fermions
with matrix elements $\big(I_{RL}^{(00)}\big)_{ij} = \delta_{ij} I_{RL}^{i(00)}$:
\begin{align}
	I_{RL}^{i(00)}
 &\equiv \dy\,f_L^if_R^i
  = \frac{\pi RM_i}{\sinh(\pi RM_i)}\ ,
\end{align}
which behaves as $2\pi RM_i\e^{-\pi RM_i}$ for $\pi RM_i \gg 1$,
thus realizing the hierarchical small quark masses without fine tuning of $M_i$.
We thus know that the matrices of Yukawa coupling
$\frac{g_4Y_u}2$ and $\frac{g_4Y_d}2$ are given as 
\begin{align} 
	\label{Yukawa coupling} 
	\frac{g_4}2Y_u = \frac{g_4}2\sqrt2I_{RL}^{(00)} U_4 \quad , \qquad
	\frac{g_4}2Y_d = \frac{g_4}2I_{RL}^{(00)} U_3
\end{align}
These matrices are diagonalized by bi-unitary transformations as in the SM
and Cabibbo-Kobayashi-Maskawa (CKM) matrix is defined in a usual way \cite{CKMmatrix}.
\begin{align}
	\label{cond_U3,U4}
	\left\{
	\begin{aligned}
	  \hat Y_d &= {\rm diag}(\hat m_d,\cdots) = V_{dR}^\dag Y_d V_{dL}\\
	  \hat Y_u &= {\rm diag}(\hat m_u,\cdots) = V_{uR}^\dag Y_u V_{uL}
	\end{aligned}
	\right.\quad , \qquad
	  V_{\rm CKM}\equiv V_{dL}^\dag V_{uL}\,
\end{align}
where all the quark masses are normalized
by the $W$-boson mass as $\hat m_f = \frac{m_f}{M_W}$.
A remarkable point is
that the Yukawa couplings $\frac{g_4Y_u}2$ and $\frac{g_4Y_d}2$ are mutually related
by the unitarity condition eq.\,\eqref{unitary_cond},
on the contrary those are completely independent in the SM.

Now physical observables $\hat m_u$, $\hat m_c$, $\hat m_d$, $\hat m_s$ 
and the Cabibbo angle $\theta_c$ are all written
in terms of $a_i$, $b_i\big(\!\equiv I_{RL}^{i(00)}$\big), $\theta$, $\theta'$ and $\gamma$.
Namely trivial relations
\begin{align} \label{qmasssimuequ}
	\hat Y_u^\dag\hat Y_u
 &= {\rm diag}\big(\hat m_u^2\,,\,\hat m_c^2\big)
	\quad , \qquad
	\hat Y_d^\dag\hat Y_d
  = {\rm diag}\big(\hat m_d^2\,,\,\hat m_s^2\big)\ .
\end{align}

\noindent
Let us note that some phases appear in ($2\times2$) CKM matrix in this parametrization.
So we change the base and eliminate any phases from ($2\times 2$) CKM here
in order to fit the phase convention of the SM.
We can achieve it by the following rephasing.
\begin{align}
	\left\{
	\begin{array}{rl}
	 u ~\to &\!\! P_uu\\[3pt]
	 d ~\to &\!\! P_dd
	\end{array}
	\right.
	\qquad\text{while}\quad
	\left\{
	\begin{array}{rl}
	   P_u &\!\!\!\!
	 = {\rm diag}\Big(\e^{i(\gamma-\theta_2)},\,\e^{i(\gamma-\theta_1)}\Big)\\[7pt]
	   P_d &\!\!\!\!
	 = {\rm diag}\Big(\e^{i(\gamma-\theta_1-\theta_2)},\,1\Big)
	\end{array}
	\right.
\end{align}
where the phases $\theta_1$ and $\theta_2$ are given as
\begin{subequations}
\begin{align}
	\tan\theta_1
 &= \frac{\sin\gamma\sin\theta_{uL}\sin\theta_{dL}}
		 {\cos\theta_{uL}\cos\theta_{dL}+\cos\gamma\sin\theta_{uL}\sin\theta_{dL}}\ ,\\
	\tan\theta_2
 &= \frac{\sin\gamma\cos\theta_{uL}\cos\theta_{dL}}
		 {\sin\theta_{uL}\cos\theta_{dL}-\cos\gamma\cos\theta_{uL}\cos\theta_{dL}}\ .
\end{align}
\end{subequations}
The $\theta_{dL}, \ \theta_{uL}$ are angles
parametrizing $V_{dL}, \ V_{uL}$, respectively:
\begin{subequations}
\begin{align} 
	\label{tanuL}
	\tan2\theta_{uL}
 &= \frac{2s_{a_1}s_{a_2}\big(b_2^2-b_1^2\big)\sin2\theta'}
    	 {\big(s_{a_1}^2-s_{a_2}^2\big)\!\big(b_1^2+b_2^2\big)
		 -\big(s_{a_1}^2+s_{a_2}^2\big)\!\big(b_2^2-b_1^2\big)\cos2\theta'}\ ,\\
	\label{tandL}
	\tan2\theta_{dL}
 &= \frac{2c_{a_1}c_{a_2}\big(b_2^2-b_1^2\big)\sin2\theta}
		 {\big(c_{a_1}^2-c_{a_2}^2\big)\!\big(b_1^2+b_2^2\big)
		 -\big(c_{a_1}^2+c_{a_2}^2\big)\!\big(b_2^2-b_1^2\big)\cos2\theta}. 
\end{align}
\end{subequations}
Then the Cabibbo angle $\theta_c$ is given as
\begin{align}
	\label{CabibboCP}
	\cos2\theta_c
 &= \cos2\theta_{uL}\cos2\theta_{dL}+\cos\gamma\sin2\theta_{uL}\sin2\theta_{dL}\ .
\end{align}
Note that 5 physical observables are written in terms of 7 parameters,
$a_i$, $b_i$ ($i = 1,2$), $\theta$, $\theta'$ and $\gamma$ in this case.
So our theory has 2 degrees of freedom, which cannot be determined by the observables.
We choose $\theta'$ and $\gamma$ as free parameters here.
Then once we choose the values of $\theta'$ and $\gamma$,
other 5 parameters can be completely fixed by the observables,
by solving eqs.\,\eqref{qmasssimuequ} and \eqref{CabibboCP} numerically
for $a_i$, $b_i$ and $\theta$.

Thus we have confirmed that observed quark masses
and flavor mixing angle can be reproduced in our model of GHU. 
Let us note that in eq.\,\eqref{CabibboCP} 
Cabibbo angle $\theta_c$ vanishes in the limit of universal bulk mass, 
{\it i.e.} $M_1 = M_2$ leads to $b_1 = b_2$ as is expected.

\section[Constraint from $\Delta S = 2$ process]{Constraint from $\bs{\Delta S = 2}$ process}
In this section,
we apply the results of the previous section to a representative CP-violating FCNC process,
$K^0$\,--\,$\bar K^0$ mixing in the down-type quark sector
caused by the non-zero KK mode gluon exchange at the tree level
as the dominant contribution to this FCNC process,
and we also estimate the lower bound on the compactification scale $R^{-1}$
from $K^0$\,--\,$\bar K^0$ mixing
responsible for the mass difference $\Delta m_K$ of two neutral $K$ mesons
and the parameter $\varepsilon_K$.%
\footnote{%
For the studies of $K^0-\bar K^0$ mixing 
in other new physics models, see for instance \cite{2001KT, CFW}.
}

We focus on the FCNC processes of zero-mode down-type quarks
due to gauge boson exchange at the tree level.
We derive the 4D effective strong interaction vertices
with respect to the zero-modes of down-type quarks relevant for our calculation:
\begin{align}
	\label{strongcCP}
	\mathcal L_\s
 \supset&\,
	\frac{g_\s}{2\sqrt{2\pi R}}G_\mu^a
	\Big(\bar d_R^i\lambda^a\gamma^\mu d_R^i+\bar d_L^i\lambda^a\gamma^\mu d_L^i\Big)
	+\sum_{n=1}^\infty\frac{g_\s}2G_\mu^{a(n)}
	 \bar d_R^i\lambda^a\gamma^\mu d_R^j\!
	 \left(P_d^\dag V_{dR}^\dag I_{RR}^{(0n0)}V_{dR}P_d\right)_{ij}\notag\\*
 &  +\sum_{n=1}^\infty\frac{g_\s}2G_\mu^{a(n)}
	\bar d_L^i\lambda^a\gamma^\mu d_L^j\!
	\left\{
	P_d^\dag V_{dL}^\dag\!
	\left(U_3^\dag I_{LL}^{(0n0)}U_3^\dag+U_4^\dag I_{LL}^{(0n0)}U_4^\dag\right)\!
	V_{dL}P_d
	\right\}_{ij}\notag\\
 \supset&~
	\frac{g_\s}{2\sqrt{2\pi R}}G_\mu^a
	\Big(\bar d_R^i\lambda^a\gamma^\mu d_R^i+\bar d_L^i\lambda^a\gamma^\mu d_L^i\Big)\notag\\*
 &	-\!\frac{g_\s\e^{i\gamma'}}4\sin2\theta_{dR}
	 \sum_{n=1}^\infty\!\left(I_{RR}^{1(0n0)}-I_{RR}^{2(0n0)}\right)\!\cdot
	 G_\mu^{a(n)}\bar s_R\lambda^a\gamma^\mu d_R\notag\\*
 &  -\!\frac{g_\s\e^{i\gamma'}}4(\alpha_d+i\beta_d)
	 \sum_{n=1}^\infty(-1)^n\!\left(I_{RR}^{1(0n0)}-I_{RR}^{2(0n0)}\right)\!\cdot
	 G_\mu^{a(n)}
	 \bar s_L\lambda^a\gamma^\mu d_L
\end{align}
where $d^i = (d, s)$.
The phase $\gamma' \equiv \gamma-(\theta_1+\theta_2)$
and the parameters $\alpha_d$ and $\beta_d$ are defined as follows:
\begin{subequations}
\begin{align} \label{alpha_d}
	\alpha_d
 \equiv&~
	{\rm Re}
	\left\{
	V_{dL}^\dagger\!
	\left(U_3^\dagger\sigma_3U_3+U_4^\dagger\sigma_3U_4\right)\!
	V_{dL}\right\}_{21}\notag\\
 =&~\frac{c_{a_1}^2+c_{a_2}^2}2\sin2\theta_{dL}\cos2\theta
	+c_{a_1}c_{a_2}\cos2\theta_{dL}\sin2\theta\notag\\*
 &  +\!\frac{s_{a_1}^2+s_{a_2}^2}2\sin2\theta_{dL}\cos2\theta'
	+s_{a_1}s_{a_2}\cos2\theta_{dL}\sin2\theta'\cos\gamma\\
	\beta_d
 \equiv&~
	{\rm Im}
	\left\{
	V_{dL}^\dagger\!
	\left(U_3^\dagger\sigma_3U_3+U_4^\dagger\sigma_3U_4\right)\!
	V_{dL}\right\}_{21}
  = -s_{a_1}s_{a_2}\sin2\theta'\sin\gamma. 
\end{align}
\end{subequations}
The $\theta_{dR}$ in the rotation matrix $V_{dR}$
to diagonalize $I_{RL}^{(00)}U_3U_3^\dag I_{RL}^{(00)}$:
\begin{align}
   \tan2\theta_{dR} 
 = \frac{2\big(c_{a_1}^2-c_{a_2}^2\big)b_1b_2\sin2\theta} 
		{\big(c_{a_1}^2+c_{a_2}^2\big)\!\big(b_1^2-b_2^2\big)
		+\big(c_{a_1}^2-c_{a_2}^2\big)\!\big(b_1^2+b_2^2\big)\cos2\theta}\ .
\end{align}
$I_{RR}^{i(0n0)}$ and $I_{LL}^{i(0n0)}$ is an overlap integral relevant to gauge interaction,
\begin{subequations} \label{vertexfunction} 
\begin{align}
	I_{RR}^{i(0n0)}
 &\equiv
	\frac1{\sqrt{\pi R}}\dy\big(f^i_R\big)^2\cos\frac nRy
  = \frac1{\sqrt{\pi R}}
	\frac{(2RM_i)^2}{(2RM_i)^2+n^2}
	\frac{(-1)^n\e^{2\pi RM_i}-1}{\e^{2\pi RM_i}-1}\ ,\\
	I_{LL}^{i(0n0)}
 &= I_{RR}^{i(0n0)}\Big|_{M_i\,\to\,-M_i}
  = (-1)^nI_{RR}^{i(0n0)}
\end{align}
\end{subequations} 
since the chirality exchange corresponds to the exchange of two fixed points. 
We can see from \eqref{strongcCP}
that the FCNC appears in the couplings of non-zero KK gluons
due to the fact that $I_{RR}^{(0n0)}$ is not proportional
to the unit matrix in the generation space,
while the coupling of the zero-mode gluon is flavor conserving, as we expected.

\begin{figure}[t]
\[
\begin{array}{ccc}
\includegraphics[bb= 0 0 106 70, scale=1]{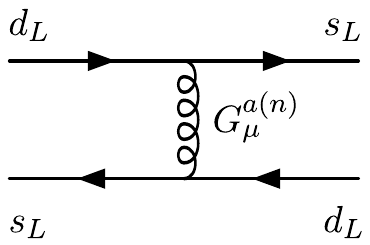} \qquad & \qquad
\includegraphics[bb= 0 0 106 70, scale=1]{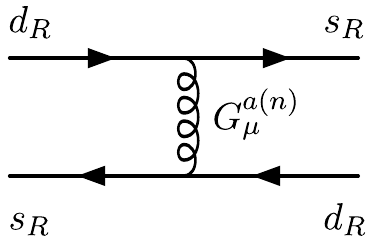} \qquad & \qquad
\includegraphics[bb= 0 0 106 70, scale=1]{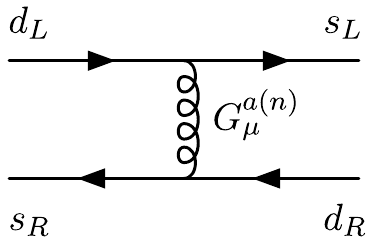}\\[8pt]
\text{(i) LL type} \qquad & \qquad
\text{(ii) RR type} \qquad & \qquad
\text{(iii) LR type}
\end{array}
\]
\caption{The diagrams of $K^0-\bar K^0$ mixing via non-zero KK gluon exchange}
\label{fig1}
\end{figure}

It turns out that the non-zero KK gluon vertex of the left-handed type
contains the CP-violating phase in any base of phase,
since it gets contributions from both of $U_3$ and $U_4$.
The non-zero KK gluon exchange diagrams,
which give the dominant contribution to the process of $K^0$\,--\,$\bar K^0$ mixing,
are depicted in \autoref{fig1}.
These diagrams are expected to give the imaginary part
in the amplitude of $K^0$\,--\,$\bar K^0$ mixing
through non-zero KK gluon exchange at the tree level,
and therefore to the parameter $\varepsilon_K$.

The contributions from each type diagram
of the $K^0$\,--\,$\bar K^0$ mixing in \autoref{fig1} is written
in the form of effective four-Fermi lagrangian, 
\begin{subequations} \label{KKbarCP}
\begin{align}
	\label{KKbarLL}
	\text{(LL type)}
 &\sim
	-\frac{\pi\alpha_\s}2
	\frac{\e^{2i\gamma'}(\alpha_d+i\beta_d)^2S_{\rm KK}^{LL}}{R^{-2}}
	\big(\bar s_L\lambda^a\gamma^\mu d_L\big)\!
	\big(\bar s_L\lambda^a\gamma_\mu d_L\big)\ ,\\
	\label{KKbarRR}
	\text{(RR type)}
 &\sim
	-\frac{\pi\alpha_\s}2\frac{\e^{2i\gamma'}\sin^2\!2\theta_{dR}S_{\rm KK}^{RR}}{R^{-2}}
	\big(\bar s_R\lambda^a\gamma^\mu d_R\big)\!
	\big(\bar s_R\lambda^a\gamma_\mu d_R\big)\ ,\\
	\label{KKbarLR}
	\text{(LR type)}
 &\sim
	-\frac{\pi\alpha_\s}2
	 \frac{\e^{2i\gamma'}(\alpha_d+i\beta_d)\sin2\theta_{dR}S_{\rm KK}^{LR}}{R^{-2}}
	 \big(\bar s_L\lambda^a\gamma^\mu d_L\big)\!
	 \big(\bar s_R\lambda^a\gamma_\mu d_R\big)
\end{align}
\end{subequations}
where four-dimensional $\alpha_\s$ is defined
by $\alpha_\s = \frac{\left(g_\s^{4D}\right)^2}{4\pi}$
with $g_\s^{4D} \equiv \frac{g_\s}{\sqrt{2\pi R}}$
and the mode sums are defined as
\begin{subequations} \label{modesumSKK}
\begin{align}
	S_{\rm KK}^{LL}
  =	S_{\rm KK}^{RR}
 &\equiv
	\pi R\sum^\infty_{n=1}\frac1{n^2}\!\left(I_{RR}^{1(0n0)}-I_{RR}^{2(0n0)}\right)^2\ ,\\
	S_{\rm KK}^{LR}
 &\equiv
	\pi R\sum^\infty_{n=1}\frac{(-1)^n}{n^2}\!
	\left(I_{RR}^{1(0n0)}-I_{RR}^{2(0n0)}\right)^2\ .
\end{align}
\end{subequations}
The sum over the integer $n$ is convergent
and the coefficients of the effective lagrangian
\eqref{KKbarCP} are suppressed by the compactification scale 
as $1/M_{\rm c}^2$ where $M_{\rm c} = R^{-1}$.

Comparing the calculation of \eqref{KKbarCP} with the experimental data,
we can obtain a lower bound on the compactification scale.
The most general effective Hamiltonian for $\Delta S = 2$ processes
due to some \lq\lq New Physics" at a high scale $\Lambda_{\rm NP} \gg M_W$
can be written as follows;
\begin{align} \label{effectiveH}
	\mathcal H_{\rm eff}^{\Delta S = 2}
  = \frac1{\Lambda_\text{NP}^2}\!
	\left(
	\sum_{i=1}^5 z_iQ_i
	+\sum_{i=1}^3 \tilde z_i\tilde Q_i
	\right)
\end{align}
where
\begin{gather}
	Q_1
  = \bar s_L^\alpha \gamma_\mu d_L^\alpha
	\bar s_L^\beta \gamma^\mu d^\beta_L\ ,\qquad
	Q_2
  = \bar s_R^\alpha d_L^\alpha\bar s_R^\beta d^\beta_L\ ,\qquad
	Q_3
  = \bar s_R^\alpha d_L^\beta\bar s_R^\beta d^\alpha_L\ ,\notag\\*
	\label{Qi}
	Q_4
  = \bar s_R^\alpha d_L^\alpha\bar s_L^\beta d^\beta_R\ ,\qquad
	Q_5
  = \bar s_R^\alpha d_L^\beta\bar s_L^\beta d^\alpha_R\ ,
\end{gather}
Indices $\alpha$, $\beta$ stand for the color degrees of freedom.
The operators $\tilde Q_{1,2,3}$ are obtained from the $Q_{1,2,3}$
by the chirality exchange $L \leftrightarrow R$.
If we assume one of these possible operators gives dominant contribution to the mixing,
each coefficient is independently constrained as follows,
with the constraints for $\tilde z_i$ are the same with those for $z_i$ ($i = 1,2,3$)
\cite{UTfit};
\begin{subequations} \label{KKconstraintsCP}
\begin{align} \label{KKconstraintsRe}
	{\rm Re}\,z_1
 &\leq [-9.6, 9.6]\times10^{-7}\,(\nicefrac{\Lambda_{\rm NP}}{1{\rm TeV}})^2\ ,\notag\\
	{\rm Re}\,z_2
 &\leq [-1.8, 1.9]\times10^{-8}\,(\nicefrac{\Lambda_{\rm NP}}{1{\rm TeV}})^2\ ,\notag\\
	{\rm Re}\,z_3
 &\leq [-6.0, 5.6]\times10^{-8}\,(\nicefrac{\Lambda_{\rm NP}}{1{\rm TeV}})^2\ ,\\
	{\rm Re}\,z_4
 &\leq [-3.6, 3.6]\times10^{-9}\,(\nicefrac{\Lambda_{\rm NP}}{1{\rm TeV}})^2\ ,\notag\\
	{\rm Re}\,z_5
 &\leq [-1.0, 1.0]\times10^{-8}\,(\nicefrac{\Lambda_{\rm NP}}{1{\rm TeV}})^2\notag
\end{align}
and
\begin{align} \label{KKconstraintsIm}
	{\rm Im}\,z_1
 &\leq [-4.4, 2.8]\times10^{-9}\,(\nicefrac{\Lambda_{\rm NP}}{1{\rm TeV}})^2\ ,\notag\\
	{\rm Im}\,z_2
 &\leq [-5.1, 9.3]\times10^{-11}\,(\nicefrac{\Lambda_{\rm NP}}{1{\rm TeV}})^2\ ,\notag\\
	{\rm Im}\,z_3
 &\leq [-3.1, 1.7]\times10^{-10}\,(\nicefrac{\Lambda_{\rm NP}}{1{\rm TeV}})^2\ ,\\
	{\rm Im}\,z_4
 &\leq [-1.8, 0.9]\times10^{-11}\,(\nicefrac{\Lambda_{\rm NP}}{1{\rm TeV}})^2\ ,\notag\\
	{\rm Im}\,z_5
 &\leq [-5.2, 2.8]\times10^{-11}\,(\nicefrac{\Lambda_{\rm NP}}{1{\rm TeV}})^2\ .\notag
\end{align}
\end{subequations}
where $\Lambda_{\rm NP}$ is regarded as the compactification scale $R^{-1}$ in our case.
All we have to do is to represent \eqref{KKbarCP}
by use of \eqref{Qi} and to utilize these constraints \eqref{KKconstraintsCP}.

We can rewrite the each type effective lagrangian \eqref{KKbarCP}
in terms of effective Hamiltonian by using the Fierz transformation 
and the completeness condition for Gell-Mann matrices;
\begin{align} \label{KKbarH}
	\mathcal H^{\Delta S=2}_{{\rm eff}, LL}
 &= \frac{z_1Q_1}{R^{-2}}\quad , \quad
	\mathcal H^{\Delta S=2}_{{\rm eff}, RR}
  = \frac{\tilde z_1\tilde Q_1}{R^{-2}}\quad , \quad
	\mathcal H^{\Delta S=2}_{{\rm eff}, LR}
  = \frac{z_4Q_4+z_5Q_5}{R^{-2}}
\end{align}
where Wilson coefficients $z_i$ are obtained as
\begin{subequations}
\begin{align}
	{\rm Re}\,z_1
 &= \frac{2\pi\alpha_\s}3
	\Big\{\big(\alpha_d^2-\beta_d^2\big)\cos2\gamma'-2\alpha_d\beta_d\sin2\gamma'\Big\}
	S_{\rm KK}^{LL}\ ,\\
	{\rm Re}\,\tilde z_1
 &= \frac{2\pi\alpha_\s}3\sin^22\theta_{dR}\cos2\gamma'S_{\rm KK}^{RR}\ ,\\
	{\rm Re}\,z_4
 &=	-2\pi\alpha_\s\sin2\theta_{dR}
	 \big(\alpha_d\cos2\gamma'-\beta_d\sin2\gamma'\big)S_{\rm KK}^{LR}\ ,\\
	{\rm Re}\,z_5 \label{KKbarzRe5}
 &= \frac{2\pi\alpha_\s}3\sin2\theta_{dR}
	\big(\alpha_d\cos2\gamma'-\beta_d\sin2\gamma'\big)S_{\rm KK}^{LR}
\end{align}
\end{subequations}
and
\begin{subequations}
\begin{align}
	{\rm Im}\,z_1
 &= \frac{2\pi\alpha_\s}3
	\Big\{\big(\alpha_d^2-\beta_d^2\big)\sin2\gamma'+2\alpha_d\beta_d\cos2\gamma'\Big\}
	S_{\rm KK}^{LL}\ ,\\
	{\rm Im}\,\tilde z_1
 &= \frac{2\pi\alpha_\s}3\sin^22\theta_{dR}\sin2\gamma'S_{\rm KK}^{RR}\ ,\\
	{\rm Im}\,z_4
 &=	-2\pi\alpha_\s\sin2\theta_{dR}
	 (\alpha_d\sin2\gamma'+\beta_d\cos2\gamma')S_{\rm KK}^{LR}\ ,\\
	{\rm Im}\,z_5 \label{KKbarzIm5}
 &= \frac{2\pi\alpha_\s}3\sin2\theta_{dR}
	(\alpha_d\sin2\gamma'+\beta_d\cos2\gamma')S_{\rm KK}^{LR}\ .
\end{align}
\end{subequations}
The constant $\alpha_\s$ should be estimated at the scale $\mu_K = 2.0$\,GeV
where the $\Delta S = 2$ process is actually measured \cite{UTfit}.
So we have to take into account the renormalization group effect
from the weak scale down to $\mu_K$
and we obtain $\alpha_\s \approx 0.268$ \cite{2010AKLM, 2011AKLM}.

Combining these results,
we obtain the lower bounds for the compactification scale $R^{-1}$
from the constraint \eqref{KKconstraintsCP}.
For that purpose,
we have to consider the \lq\lq weight" for the contribution of each type diagram
from lattice QCD calculations of the matrix elements
with non-perturbative renormalization.
As is discussed precisely  in the ref \cite{UTfit},
it is known that
an analytic formula for the contribution to the $K^0$\,--\,$\bar K^0$ mixing amplitudes
induced by a given New Physics scale coefficient,
denoted by $\big\langle\bar K^0\big|\mathcal H_{\rm eff}\big|K^0\big\rangle_i$\,:
\begin{align}
	\big\langle\bar K^0\big|\mathcal H_{\rm eff}\big|K^0\big\rangle_i
 &= \sum_{j=1}^5\sum_{r=1}^5
	\left(b_j^{(r,i)}+\eta c_j^{(r,i)}\right)\!\eta^{a_j}
	\frac{z_i(\Lambda_{\rm NP})}{\Lambda_{\rm NP}^2}R_r
	\big\langle\bar K^0\big|Q_1\big|K^0\big\rangle\ ,
\end{align}
where $\eta \equiv \alpha_\s(\Lambda_{\rm NP})/\alpha_\s(m_t)$ ($m_t$ : top quark mass),
the \lq\lq magic numbers" $a_j$, $b_j^{(r,i)}$ and $c_j^{(r,i)}$ can be found
in ref. \cite{1998Ciuchini}.
$R_r$ are the ratios of the matrix elements of the New Physics operators $Q_i$
over the SM one \cite{UTfit}:
\begin{align}
	R_1 = 1 \quad, \quad
	R_2 \simeq -12.9 \quad , \quad
	R_3 \simeq 3.98 \quad , \quad
	R_4 \simeq 20.8 \quad , \quad
	R_5 \simeq 5.2\ .
\end{align}
Utilizing these values, the contribution of LR type diagram ($z_4$ and $z_5$) can be written
as just like that of LL type ($z_1$);
\begin{align}
	\mathcal H_{{\rm eff}, LL+RR+LR}^{\Delta S = 2}
 &= \frac{\mathcal Z_1Q_1}{R^{-2}}
	\qquad\text{where}\quad
	\mathcal Z_1
  \simeq 
	z_1+\tilde z_1-294\cdot z_4-97.2\cdot z_5
\end{align}
Note that we use the fact
$\big\langle\bar K^0\big|Q_1\big|K^0\big\rangle
= \big\langle\bar K^0\big|\tilde Q_1\big|K^0\big\rangle$
due to the parity symmetry of strong interaction.
Then we can get lower bound on the compactification scale $R^{-1}$ by use of the upper bound
on the relevant coefficients $z_1$ given in \eqref{KKconstraintsCP}:
\begin{subequations}
\begin{align}
	\text{Real part : }~\frac1R
 &\gtrsim
	\sqrt{\frac{|{\rm Re}\,\mathcal Z_1|}{9.6\times10^{-7}}}~\big[{\rm TeV}\big]\ ,\\
	\text{Imaginary part : }~\frac1R
 &\gtrsim
	\left\{
	\begin{array}{lc}
	  \dis\!\!\sqrt{\frac{{\rm Im}\,\mathcal Z_1}{2.8\times10^{-9}}}~\big[{\rm TeV}\big]
	&\quad \text{for }~{\rm Im}\,\mathcal Z_1 > 0\\[10pt]
	 \dis\!\!\sqrt{-\frac{{\rm Im}\,\mathcal Z_1}{4.4\times10^{-9}}}~\big[{\rm TeV}\big]
	&\quad \text{for }~{\rm Im}\,\mathcal Z_1 < 0
	\end{array}
	\right.\ .
\end{align}
\end{subequations}

\begin{figure}[!t]
\centering
	\includegraphics[bb=0 00 600 440, scale=0.5]{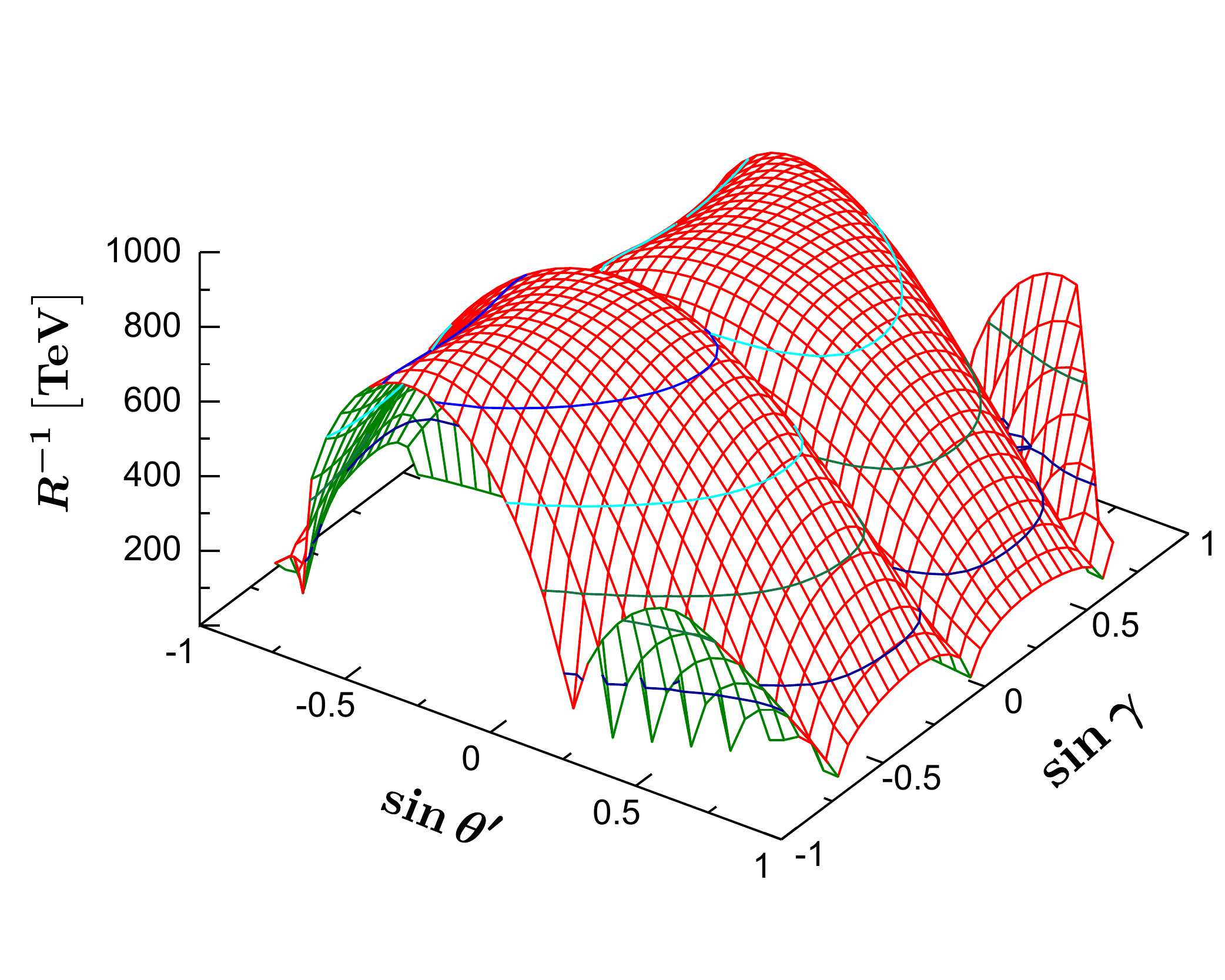}
	\\[-10pt]
\caption{\small%
The lower bound for $R^{-1}$ from $\Delta S = 2$ process.
}
\label{fig:CPbound}
\end{figure}

\noindent
We can get the lower bound on $R^{-1}$ from $\Delta S = 2$ process in our model
by combining these constraints.
Since our theory has two free parameters, say $\theta'$ and $\gamma$,
the lower bound on $R^{-1}$ depends on it.
The obtained numerical result is displayed in \autoref{fig:CPbound},
where the lower bound on $R^{-1}$ is plotted as a function of ($\sin\theta'$, $\sin\gamma$). 
The phenomenological constraint on $R^{-1}$ is from a few TeV to about 900\,TeV
from this result.


\section{Summary}
In this paper,
we discussed the Kobayashi-Maskawa type CP violation
in the context of 5D $SU(3)\otimes SU(3)_{\rm color}$ gauge-Higgs unification.
In this model, a pair of $\bf 3$ and $\bar{\bf 6}$ representation of $SU(3)$ should be introduced to reproduce a generation, 
and then, an extra quark doublet per generation appears.
So we identify the quark doublets corresponding Standard Model $Q_{\rm SM}$
as some combination of them.
These combinations generically contains complex components,
it potentially violates CP symmetry.

As was mentioned in introduction, the $n$ generation model is considered 
and the Yukawa coupling has $(n-1)^2$ phases after re-phasing.
Then the similar discussion of KM theory can be proceeded
and CP symmetry breaks down more than the three generation at zero-mode sector
as is in the SM.
On the other hand, the FCNC vertices in the strong interaction
between the zero-mode fermion and non-zero KK mode gluon exist in this model 
and thus CP-violating phase appears in the strong interaction.

In fact, one can show that the CP-violating phases disappear 
when the flavor symmetry is restored, which corresponds to the case of degenerate fermion bulk mass term.
In the main text, we construct the two generation model and it can be easily extended to the $n$ generation model.
In this case,
the CP-violating phases included in the coefficients of FCNC vertices in the strong interaction \eqref{strongcCP} vanishes
in the limit of universal bulk masses $M_1 = M_2 = \cdots$
by use of the unitarity condition \eqref{unitary_cond}:
\begin{subequations}
\begin{align}
	P_d^\dag V_{dR}^\dag I_{RR}^{(0m0)}V_{dR}P_d
 &~\xrightarrow{M_1 = M_2 =\, \cdots\,}~
	P_d^\dag V_{dR}^\dag V_{dR}P_dI_{RR}^{(0m0)}
  = I_{RR}^{(0m0)}
\end{align}
and
\begin{align}
 &  P_d^\dag
	V_{dL}^\dag\!
	\left(
	U_3^\dag I_{RR}^{(0m0)}U_3
	+U_4^\dag I_{RR}^{(0m0)}U_4
	\right)\!
	V_{dL}P_d\notag\\
 \xrightarrow{M_1 = M_2 =\, \cdots\,}~&
	P_d^\dag V_{dL}^\dag\!
	\left(U_3^\dag U_3+U_4^\dag U_4\right)\!
	V_{dL}P_dI_{RR}^{(0m0)}
 =I_{RR}^{(0m0)}\ .
\end{align}
\end{subequations}
Then, one can see that the CP violation is raised together with  flavor violation.  

{}From the above argument, the CP violation generally occurs in the even two generation,
we construct the two generation model as a simplest example of CP-violating model.
Thereupon, the neutral kaon system is discussed in this context
and the rate of $K^0 -\bar K^0$ mixing is estimated.
We calculate the Wilson coefficient caused by the $\Delta S = 2$ process
and obtain the lower bound of the compactification scale $R^{-1}$
by comparing the mass difference $\Delta m_K$ and $\varepsilon_K$ which is known as a typical CP-violating observables.
The LL type diagrams give stringent constraint in most cases, 
we put the phenomenological constraint  on $R^{-1}$ as $\mathcal O(1)$ TeV $\sim \mathcal O(10^3)$ TeV.

\subsection*{Acknowledgments}
We would like to thank Professor C. S. Lim
for fruitful discussions and valuable suggestions at various stages of this work.
This work was supported in part by the Grant-in-Aid for Scientific Research
of the Ministry of Education, Science and Culture, No.~21244036 and 
in part by Keio Gijuku Academic Development Funds (N.M.).


\end{document}